\documentstyle[aps,prl,twocolumn,psfig]{revtex}
\begin{document}
\draft

\title{Current noise, electron-electron interactions, and quantum 
interference}
\author{Felix von Oppen and Ady Stern}
\address{Department of Condensed Matter Physics, Weizmann Institute of 
Science, 76100 Rehovot, Israel}
\date{\today}

\twocolumn[
\maketitle
\widetext%
\vspace*{-6.45mm}
\leftskip=1.9cm\rightskip=1.9cm
\begin{abstract}
  It is  shown that electron-electron   interactions lead to  a  novel
  quantum-interference contribution  to  non-equilibrium current noise
  in  mesoscopic  conductors.  The   corresponding noise   spectrum is
  obtained in  detail for diffusive systems and  found to be quadratic
  in the applied voltage and to exhibit a  power-law dependence on the
  frequency  $S_{\rm ee}(\omega)\sim|\omega|^{-\alpha}$.  The exponent
  $\alpha=  (4-d)/2$ depends on the dimensionality  $d$ of the sample. 
  \vspace*{-1mm}
\noindent
\end{abstract}
\vspace{0mm}
]

\narrowtext 

Various sources  of current noise in  mesoscopic conductors  have been
investigated                in             recent                years
\cite{Lesovik,Jong,Reznikov,Feng,Landauer}.   In  thermal equilibrium,
the    noise  power  is   related   to     the  conductance  via   the
fluctuation-dissipation theorem.   Such  a general  relation  does not
hold at finite bias  voltages in which case  the noise power  contains
information beyond the  conductance.  For example, the quantization of
the   electronic charge manifests   itself  in  shot  noise at  finite
voltages. In mesoscopic  conductors, shot noise is suppressed relative
to  its classical value   due  to Fermi correlations  \cite{Lesovik}.  
Quantum coherence leads   to    a weak-localization  correction    and
mesoscopic  fluctuations    of  the   shot-noise  power   \cite{Jong}. 
Sensitive  measurements of noise  in mesoscopic  samples have recently
become feasible and some predictions on shot noise have been confirmed
experimentally \cite{Reznikov}.  Besides these noise sources intrinsic
to  the  electron system, the coupling   to  non-electronic degrees of
freedom such as  fluctuating lattice impurities is  believed to be the
origin of $1/f$ noise \cite{Feng}.

It is the purpose of the  present paper to show that electron-electron
interactions  lead  to  a   novel   quantum-coherence contribution  to
non-equilibrium noise in mesoscopic  systems.  This source of  current
noise is {\it   intrinsic} to the   electron system and  distinct from
equilibrium  and shot noise.   Its origin is the electromagnetic field
experienced by an electron moving in a mesoscopic sample.  This field,
which  mediates  the  interaction between   the  electrons, fluctuates
because  of the thermal  fluctuations    in  the charge and    current
densities in  the system.  In a  seminal paper, Altshuler, Aronov, and
Khmelnitsky (AAK) \cite{AAK} have studied how these field fluctuations
suppress  the weak-localization   correction to the  impurity-averaged
conductivity.  Here  we  show  that the  fluctuating field   makes the
sample-specific conductivity time dependent  and thus produces current
noise.

For a  qualitative  picture of the   effect, we  consider the  current
response to a uniform {\it dc} electric field ${\bf E}_{\rm dc}$,
\begin{equation}
  {\bf j}({\bf r},t)=\int_{-\infty}^t dt^\prime\int d{\bf r^\prime}\,
      \sigma({\bf r},t;{\bf r^\prime},t^\prime){\bf E}_{\rm dc}.
\end{equation}
The  sample-specific conductivity $\sigma$  depends explicitly  on two
space-time points  because  the fluctuating field breaks  both spatial
and  temporal     translation  invariance.   In    the   semiclassical
approximation the conductivity can  be written  as  a double sum  over
classical paths $\alpha$ and  $\beta$ from $({\bf r^\prime},t^\prime)$
to $({\bf  r},t)$, namely  
\begin{equation}
   \sigma({\bf r},t;{\bf  r^\prime},t^\prime)\sim\sum_{\alpha,\beta}
      A_\alpha^*A_\beta\exp[i(S_\alpha-S_\beta)/\hbar].
\end{equation}  
Here $A_\alpha$  denotes  a classical  probability  amplitude of  path
$\alpha$  and  the phase   factor is   determined  by  the  associated
classical action $S_\alpha$.   Quantum corrections to the conductivity
arise  from the  interference of  distinct  paths $\alpha\ne\beta$ for
which the phases do not cancel out.  In the absence of the fluctuating
field, the phases accumulated along the two diffusion paths due to the
static disorder potential    give  rise to   a  random  {\it   static}
interference pattern. In this  way, sample-to-sample variations in the
electrostatic disorder potential   lead to the universal   conductance
fluctuations \cite{Lee}.  {\it The fluctuations of the electromagnetic
  field lead, in addition, to a  time-dependent potential landscape of
  a  particular  sample,   thus  making  also the   conductivity  time
  dependent.} Specifically, the fluctuating field leaves the classical
paths essentially unchanged but  adds time-dependent  contributions to
the phases.   When measuring the {\it  dc} conductance,  the resulting
time-dependent interference  pattern  is effectively  averaged out and
the  {\it  dc}   conductance   fluctuations  are   suppressed  by  the
fluctuating field.  In  this  sense, the  fluctuating field  leads  to
dephasing.    Nevertheless,  when   measuring  current    noise,   the
interference   is    still   effective   because  the   time-dependent
interference pattern     results in  temporal   fluctuations   of  the
conductivity.

The noise amplitude is characterized by the power spectrum
\begin{equation}
   S(\omega)={1\over2}\int d\zeta\,e^{i\omega\zeta}
      \langle\{\delta{\hat I}(t),\delta{\hat I}(t+\zeta)\}\rangle_T,
\label{power}
\end{equation}
where $\langle\ldots\rangle_T$ is the thermal expectation value in the
presence  of the bias  voltage, $\delta{\hat I}={\hat I}-{\hat I}_{\rm
  dc}$, and  the curly brackets  denote the anticommutator.  According
to the  qualitative considerations given  above, the  largest possible
equal-time  current fluctuations  $\langle \delta   I^2(t)\rangle=\int
d\omega\,S_{\rm ee}(\omega)$   due   to  the    electromagnetic  field
correspond to a situation when the entire interference pattern becomes
time dependent.  In this situation,  the current fluctuations would be
of order $\langle  \delta G^2\rangle V^2$,  in terms  of the universal
conductance fluctuations  $\langle \delta G^2\rangle$ and  the applied
voltage $V$.    We find for  one-dimensional  systems that the current
fluctuations do indeed reach this upper bound when the phase-coherence
time $\tau_\phi$  is comparable to   the diffusion time $\tau_D=L^2/D$
through the sample  ($L$  is the sample  size  and  $D$ the  diffusion
constant).  We  also show that the  same  is no longer true  in higher
dimensions.

The frequency  spectrum  of the  noise is  sensitive  to  the specific
nature  of the  fluctuating  field.  Under  realistic conditions,  the
field is essentially electric and may be characterized in terms of the
correlator  of  the      scalar  potential   $\varphi$.    Using   the
fluctuation-dissipation  theorem,  one      obtains  for   frequencies
$\hbar\omega\ll T$ \cite{AAK}
\begin{equation}
    \overline{\varphi({\bf k},\omega)\varphi(-{\bf k},-\omega)}
                       ={2T\over\sigma k^2}.
\label{field-correlator}
\end{equation}
Phase-sensitive quantities are predominantly affected by the classical
modes of the   fluctuating field with  $\hbar\omega\ll T$.   This is a
consequence of  the infrared divergence  in (\ref{field-correlator}).  
In particular, AAK showed that the  dephasing rate $\tau_\phi^{-1}$ is
determined  by these modes.    By contrast, the energy-relaxation rate
$\tau_{\rm  ee}^{-1}$ is determined  by modes with $\hbar\omega\sim T$
\cite{AAK,Schmid}.  In    one  and two   dimensions,  AAK   found that
$\tau_\phi\ll\tau_{\rm ee}$  while in  three dimensions the  two times
are  of the same  order.  Eq.\ (\ref{field-correlator}) shows that the
temporal correlations  of  $\varphi({\bf r},t)$ are short  ranged with
characteristic  scale $\hbar/T$.   On  the  other  hand, the  infrared
divergence of  the    correlator (\ref{field-correlator})   emphasizes
long-range  spatial  correlations.  This  generates a second frequency
scale much  smaller  than the  temperature $T$:   the inverse time  of
flight $\tau^{-1}_D$ along  a  typical diffusion path  contributing to
the  time-dependent    interference pattern.     We   find  that   the
characteristic  frequencies  $\tau_D^{-1}$   and  $T/\hbar$  enter  as
cutoffs       on         a      power-law     frequency       spectrum
$S_{\rm ee}(\omega)\sim|\omega|^{-(4-d)/2}$.

In this paper we follow AAK and include electron-electron interactions
by averaging over   the fluctuating field  within the impurity-diagram
technique. It is also instructive to  interpret the noise contribution
discussed here  in    terms    of   many-body  diagrams    using    an
exact-eigenstate basis of the impurity potential.  In the latter case,
a typical diagram contributing  to the noise power  is shown  in Fig.\ 
1a.  Averaging over the fluctuating  field corresponds to an  RPA-like
treatment   of the electron-electron  interactions.  This diagrammatic
approach  also clarifies the  distinction between  shot  noise and the
noise discussed  in this paper.   In the limit  $eV<T$, shot  noise is
represented by diagrams in which the four external legs emanate from a
single electron loop.

We  now proceed to sketch  the  calculation of the interaction-induced
noise   power  for  diffusive conductors   \cite{details}.   Including
electron-electron  interactions   by  a  (Gaussian)   average over the
fluctuating field, the   general expression for the  disorder-averaged
noise power (\ref{power}) reduces 

\centerline{\psfig{figure=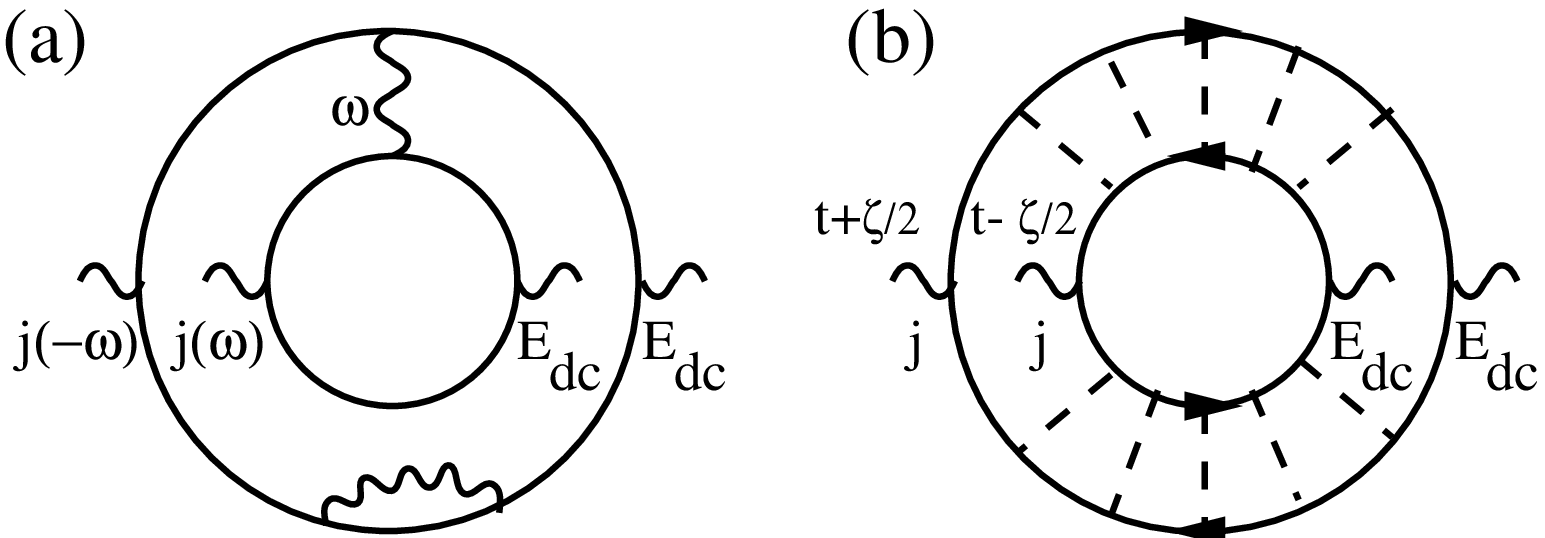,width=7.3cm}}

\begin{figure}
\caption{Diagrams for the noise power. In (a) we show a representative 
  many-body  diagram in  the  exact-eigenstate  representation  of the
  impurity potential. In (b) we explicitly  averaged over the impurity
  potential     to    leading order  in       the diffusive regime and
  electron-electron  interactions are  included by averaging  over the
  thermal fluctuations of the electric   field  in the sample.    Full
  lines  represent    electron propagators,    wavy    lines   Coulomb
  interactions, and dashed lines impurity scattering.}
\end{figure}

\noindent to
\begin{equation}
  S_{\rm ee}(\omega)=\int d\zeta\,e^{i\omega \zeta} \left\langle\overline{
    I(t+\zeta)\,I(t)}-\overline{I(t+\zeta)}\,\overline{I(t)}\right
    \rangle.
\label{definition}
\end{equation}
The overbar denotes the average over the fluctuating field and angular
brackets the disorder average.  We compute $S_{\rm ee}(\omega)$ by the
impurity-diagram technique, including the applied  {\it dc} voltage to
quadratic order.  In accordance  with the qualitative  discussion, the
relevant diagram shown in Fig.\ 1b is that responsible for conductance
fluctuations   \cite{Lee}.  This   diagram   represents  the   current
correlations for  a  particular realization of the  fluctuating field,
which  eventually needs to  be averaged   over,  as specified  in Eq.\ 
(\ref{definition}).   While there are additional diagrams contributing
to the conductance  fluctuations,  it turns  out that only  the one in
Fig.\ 1b contributes to the noise power \cite{details}.  We employ the
Keldysh technique \cite{Rammer} to evaluate this diagram in space-time
representation in the presence of a fluctuating  electric field in the
region  $L_T\ll L$.  This is  the  most interesting region because the
effect is largest for  $\tau_D\simeq\tau_\phi$ and the thermal  length
$L_T=\sqrt{h    D/T}$  of metals  is  usually   much  smaller than the
phase-coherence length $L_\phi$ \cite{AAK}. We find
\begin{eqnarray}
    \langle j(t+&&\zeta/2)j(t-\zeta/2)\rangle={4\pi\over3}\left(e^2\over
     h\right)^2{\hbar/\tau_D\over T}\,{D\over L^{2d-2}}
     \nonumber\\
     &&\times\int d{\bf r}\,d{\bf r^\prime}\int_{-\infty}^t 
     dt^\prime\,P^\zeta_{t,t^\prime}({\bf r},{\bf r^\prime}) 
     P^{-\zeta}_{t,t^\prime}({\bf r},{\bf r^\prime}) E_{\rm dc}^2
\label{jthroughp}
\end{eqnarray}
in  terms  of  the diffuson  propagator $P^\zeta_{t,t^\prime}({\bf r},
{\bf r^\prime})$ in the  presence of the  electric potential.  In  the
following, we  will restrict attention  to  the region $L_T<L< L_\phi$
where we   can neglect  the  effects of  processes  with large  energy
transfers $\hbar\omega\sim T$.

Diagrammatically,    the diffuson  $P_{t,t^\prime}^\zeta({\bf  r},{\bf
  r^\prime})$  is represented by Fig.\ 2,  and can be shown to satisfy
the differential equation
\begin{eqnarray}
   \left\{\partial_t+{ie\over\hbar}\varphi_\zeta({\bf r},t)+D\nabla^2
    \right\}&&P_{t,t^\prime}^\zeta({\bf r},{\bf r^\prime})
     \nonumber\\
     &&=\delta(t-t^\prime)
     \delta({\bf r}-{\bf r^\prime}).
\label{diffequation}
\end{eqnarray}

\centerline{\psfig{figure=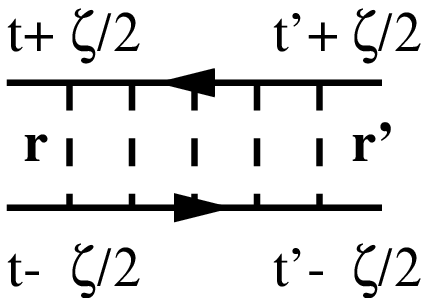,width=3cm}}

\begin{figure}
\caption{Diagrammatic representation of the diffuson in real 
  space and time. Full and dashed lines as in Fig.\ 1b.}
\end{figure}

\noindent In the absence  of time-dependent fields,  the phase 
accumulated by  an electron along a  diffusive path is  independent of
when  the trajectory is traversed.   In this  case,  the phases of the
advanced and  retarded   Green functions entering  into  the  diffuson
diagram in Fig.\ 2 cancel and the  diffuson is independent of the time
difference $\zeta$.  By contrast,   the accumulated phase due  to  the
fluctuating field does depend on  when the path  is traversed and  for
this reason, the scalar potential enters the equation for the diffuson
in the combination
\begin{equation}
    \varphi_\zeta({\bf r},t)=\varphi_A({\bf r},t+\zeta/2)-
                   \varphi_R({\bf r},t-\zeta/2).
\end{equation}
Here we have allowed  for   the  possibility  that the  advanced   and
retarded Green functions need to be evaluated for different potentials
$\varphi_A({\bf r},t)$  and  $\varphi_R({\bf r},t)$,  respectively, as
required  for the factorized average  in (\ref{definition}).  That the
fluctuating field affects the diffuson propagator (\ref{diffequation})
only via   time-dependent  phases is  particularly  evident  from  the
path-integral solution for the diffuson (\ref{diffequation})
\begin{eqnarray}    
   &&P_{t,t^\prime}^\zeta({\bf r},{\bf r^\prime})=\int_{{\bf r}
   (t^\prime)={\bf r^\prime}}^{{\bf r}(t)={\bf r}}[d{\bf r}(\tau)]
   \nonumber\\
   &&\,\,\,\,\,\,\,\,\,\,\times\exp\left\{-\int_{t^\prime}^t d\tau\left[
   {{\dot{\bf r}}^2(\tau)\over 4D}+{ie\over\hbar}
   \varphi_\zeta({\bf r}(\tau),\tau)\right]\right\}.   
\end{eqnarray}
One observes that the contribution of a path is determined not only by
its diffusive weight, but also by the difference of the time-dependent
phases accumulated by the advanced and retarded Green functions.

To   perform the (Gaussian)   average  over the field fluctuation,  we
collect  the phases  associated with  the   two diffusive paths  ${\bf
  r}_1(\tau)$ and ${\bf r}_2(\tau)$ entering Eq.~(\ref{jthroughp}).
The total  phase $\phi=\phi_++\phi_-$ is the  sum of two contributions
associated with the two conductivity bubbles,
\begin{eqnarray}
 \phi_\pm&=&\pm{e\over\hbar}\int_{t^\prime}^t d\tau\,\left[
    \varphi({\bf r}_1(\tau),\tau\pm\zeta/2)\right.
   \nonumber\\
   &&\,\,\,\,\,\,\,\,\,\,\,\,\,\,\,\,\,\,\,\,\,\,\,\,\,\,\,\,\,\,\,\,\,\,
    \,\,\,\,\,\,\,\,\left.
   -\varphi({\bf r}_2(\tau),\tau\pm\zeta/2)\right].
   \label{phi}
\end{eqnarray}
For the factorized   average in Eq.~(\ref{definition}),   one averages
over $\phi_+$    and $\phi_-$ independently,  while   the unfactorized
average includes the correlations between $\phi_+$ and $\phi_-$. Hence
\begin{eqnarray}
   \overline{\exp[i(\phi_++\phi_-)]}&&-\overline{\exp[i\phi_+]}\,
     \overline{\exp[i\phi_-]}
     \nonumber\\
     &&=\exp[-\overline{\phi_+^2}]
     \left\{\exp[-\overline{\phi_+\phi_-}]-1\right\}
     \nonumber\\
     &&\simeq-\overline{\phi_+\phi_-}.
\label{phase-average}
\end{eqnarray}
The last line  holds because the phases  due to the  fluctuating field
are  small  for $L_\phi>L$.  The  correlator $\overline{\phi_+\phi_-}$
can  be   computed         using  the   potential         fluctuations
(\ref{field-correlator}).   Due   to   the infrared   divergence    in
(\ref{field-correlator}),  it  decays  as   a  function  of the   time
difference  $\zeta$ on the  scale of the  diffusion time $\tau_D$.  We
note that  for large systems with  $L>L_\phi$ one  may also obtain the
suppression of     the   {\it dc} conductance   fluctuations    by the
fluctuating field from the factorized average in (\ref{phase-average})
\cite{AAK}.  In this way,  one naturally recovers the expressions  for
the phase-coherence time $\tau_\phi$  derived by AAK. In one dimension
one  finds $\tau_\phi^{-1}=   (4e^2\sqrt{2D}T/3\sqrt{\pi}\hbar^2\sigma
)^{2/3}$, in two  dimensions  $\tau^{-1}_\phi=(e^2T/2\pi\sigma\hbar^2)
\ln(T\tau_\phi/\hbar)$,      and       in          three    dimensions
$\tau_\phi^{-1}=e^2T^{3/2}/2\pi\sigma\hbar^{5/2}\sqrt{6D}$.

With these ingredients, we  obtain for the  noise power in the  regime
$L_T<L<L_\phi$
\begin{equation}
   S_{\rm ee}(\omega)=\left\{\begin{array}{lr}
      c_1\left(\textstyle e^2\over\textstyle h\right)^2
      \left(\textstyle L_T\over\textstyle L\right)^2
      {\textstyle \tau_D V^2\over\textstyle(|\omega|\tau_\phi)^{3/2}}&
      d=1\\
      c_2\left(\textstyle e^2\over\textstyle h\right)^2\left(\textstyle L_T
      \over\textstyle L_\phi\right)^2
      {\textstyle V^2\over\textstyle|\omega|\ln(T\tau_\phi/\hbar)}&
      d=2\\
      c_3\left(\textstyle e^2\over\textstyle h\right)^2
      \left(\textstyle L_T\over\textstyle L_\phi
      \right)^2{\textstyle V^2\over\textstyle(T|\omega|/\hbar)^{1/2}}&
       d=3\end{array}\right.
\label{theresult}
\end{equation}
The numerical  prefactors   are $c_1\simeq0.01$,  $c_2\simeq0.03$, and
$c_3\simeq0.06$.   We  also  remark  that in   two dimensions  $S_{\rm
  ee}(\omega)$  changes only by  a factor $(L_\phi/L)^2$ in the regime
$L_\phi<L<L_{\rm   ee}$.   These   results  are   valid  for  voltages
$eV<\hbar/\tau_D$   \cite{Larkin}.   The  power-law frequency  spectra
extend over the interval $1/\tau_D<|\omega|<T/\hbar$.  Outside of this
interval, $S_{\rm  ee}(\omega)$ saturates for $\hbar|\omega|\ll T$ and
vanishes  for $\hbar|\omega|\gg   T$.    A   sample  has  a    reduced
dimensionality  if its transverse dimensions  are  smaller than $L_T$. 
The dependence  of the noise  power on dimensionality  $d$ arises from
two sources.  The   dimensional dependence of diffusion  predominantly
affects    the  prefactor, analogous   to   the universal  conductance
fluctuations \cite{Lee}.  The  characteristic dependence  of the noise
spectrum on the dimensionality  $d$ originates from the $d$ dependence
of     the    degree   of    divergence     of   the  field correlator
(\ref{field-correlator}).     Effectively,  long-range correlations in
space are emphasized   less in higher  dimensions.   This  leads to  a
faster   decrease   of  $\overline{\phi_+\phi_-}$   with   $\zeta$ and
consequently to a weaker emphasis on small frequencies.

An important feature of the results in the coherent regime is that the
noise power  $S_{\rm ee}(\omega)$ does {\it  not} depend explicitly on
the  temperature.  Instead,  the  temperature   enters only into   the
frequency   scales and thereby  determines   the frequency range where
these power laws can be  observed.  This insensitivity of the spectrum
to the temperature is  a result of  the competition between the linear
increase  with  temperature of  the  field fluctuations  and the $1/T$
prefactor in (\ref{jthroughp}).  Of course, for fixed sample size $L$,
the range over  which $S_{\rm ee}(\omega)$  is temperature independent
is limited by the condition $L_T<L<L_\phi$.

It  is  instructive  to   compare these results   to the   conductance
fluctuations \cite{Lee}  $\langle\delta   G^2\rangle\sim(e^2/h)^2(L_T/
L)^2f_d(L/L_T)$ with $f_1(x)=1$, $f_2(x)=\ln  x$, and  $f_3(x)=x$.  In
the introduction we  argued that $\langle\delta  G^2\rangle V^2$ is an
upper    bound  for     the   current    fluctations    $\langle\delta
I^2(t)\rangle=\int d\omega  S_{\rm ee}(\omega)$.   In one dimension we
find          from   Eq.\    (\ref{theresult})  that    $\langle\delta
I^2(t)\rangle\sim(e^2/h)^2(L_T/L)^2(\tau_D/ \tau_\phi)^{3/2}   V^2$.   
This shows  that  the current  fluctuations do  indeed reach the upper
limit   for $\tau_D\simeq\tau_\phi$.   For    $\tau_D<\tau_\phi$,  the
reduction factor $(\tau_D/\tau_\phi)^{3/2}$  is just  the magnitude of
the  time-dependent phase accumulated  along  a typical path of length
$\tau_D$ \cite{AAK}.   By  contrast, the  current  fluctuations do not
reach the  upper bound in higher  dimensions.   In two dimensions, one
finds  $\langle\delta     I^2(t)\rangle   \sim (e^2/h)^2(L_T/L_\phi)^2
(\ln(T\tau_D/\hbar)/\ln(T\tau_\phi/\hbar))V^2$.          Even      for
$\tau_\phi=\tau_D$ the current fluctuations are smaller than the upper
bound by a  logarithmic  factor.  This is due   to the fact  that  the
weight of short paths in the noise power is  reduced relative to their
weight  in the  conductance fluctuations  because  the  time-dependent
phase  accumulated by  trajectories  much shorter than  $\tau_\phi$ is
small compared to one.  The reduction is significant in two dimensions
because  of  the logarithmic   contribution of  short  trajectories of
length $\hbar/T$ to the conductance fluctuations \cite{Lee}.  In three
dimensions,  the conductance  fluctuations  are entirely  due to short
trajectories   \cite{Lee}.   Therefore,   the     current fluctuations
$\langle\delta  I^2(t)\rangle \sim  (e^2/h)^2 (L_T/L_\phi)^2V^2$   are
strongly   reduced --     by  a   factor $(\hbar/T\tau_D)^{1/2}$   for
$\tau_D=\tau_\phi$ -- relative to the upper bound.

We conclude  by comparing   the  interaction-induced noise   to  other
sources of noise.  The  origin   of the $1/f$-like  frequency  spectra
found here  is  very different from the  mechanism  for standard $1/f$
noise.  In the usual argument for $1/f$ noise \cite{Feng}, both in the
absence and the presence of quantum coherence, the fluctuations of the
impurity  potential    are supposed    to  have  short-range   spatial
correlations and to  vary slowly in time  compared to  electronic time
scales.  The $1/f$ noise    spectrum arises because of  the  activated
nature  of impurity  motion and  the  broad distribution of activation
energies.  By contrast, for   the interaction-induced noise  the field
fluctuations are faster  than the relevant  electronic time scales but
have long-range spatial correlations. The $1/f$-like frequency spectra
reflect   the    power-law   divergence   of  the  field    correlator
(\ref{field-correlator}).   Due to  this  difference, the lower cutoff
frequency is vastly larger  for the interaction-induced noise than for
$1/f$  noise.  Moreover,   $1/f$  noise  in the  presence   of quantum
coherence  is sensitive to weak magnetic  fields since the conductance
fluctuations decrease by   a factor of  four when  applying a magnetic
field.  An  analogous    effect   does  {\it   not} occur  for     the
interaction-induced noise power \cite{details}.

The interaction-induced noise can be distinguished experimentally from
equilibrium noise by its voltage dependence.  Shot noise in the regime
$eV<T$,  on the other  hand, has  the same  voltage  dependence as the
interaction-induced   noise, but   its  frequency  spectrum, which  is
frequency   independent for    $\hbar\omega<T$, is   different.    The
shot-noise magnitude is reduced compared to its zero-temperature value
by a factor  $eV/T$, $S_{\rm shot}\sim(e^2/h)g[(eV)^2/T]$ with $g$ the
dimensionless conductance  \cite{Landauer}.  As  discussed by  de Jong
and  Beenakker  \cite{Jong}, quantum-coherence contributions  to  shot
noise are suppressed by $1/g$.  By comparison, the interaction-induced
noise  is largest  in  one  dimension for  $\tau_D\simeq\tau_\phi$ and
frequency $\omega\sim 1/\tau_D$.  From  Eq.\ (\ref{theresult}) one has
for      the         corresponding         noise      power    $S_{\rm
  ee}(\omega\sim1/\tau_D)\sim(e^2/\hbar) [(eV)^2 /T]$.  This is of the
same order as the quantum-coherence  corrections to shot noise.  These
estimates  show  that  the  interaction-induced noise  is particularly
important    in  almost localized   systems with   small dimensionless
conductance $g\sim1$.

We benefitted from discussions with Y.\ Gefen, Y.\ Imry, S.\ Tomsovic,
and    H.A.\ Weidenm\"uller.  FvO   is  supported by an Amos-de-Shalit
scholarship  of   the   Minerva   Foundation,  and  acknowledges   the
hospitality   and  support of   the  ITP  Santa   Barbara  (NSF  grant
PHY94-07194) and the MPI Dresden where some of this work was done.  AS
thanks the US-Israel  Binational Science Foundation (95-250/1) and the
Minerva Foundation for financial support.


\begin{references}

\bibitem{Lesovik}  G.B.\ Lesovik, JETP  Lett.   {\bf 49}, 592  (1989);
  C.W.J.\ Beenakker and M. B\"uttiker,  Phys.\ Rev.\ B {\bf 46},  1889
  (1992).
  
\bibitem{Jong}  M.J.M.\ de Jong  and  C.W.J.\ Beenakker, Phys.\ Rev. B
  {\bf 46}, 13400 (1992).
  
\bibitem{Reznikov} M.\  Reznikov, M.\ Heiblum,  H.\ Shtrikman, and D.\
  Mahalu, Phys.\ Rev.\  Lett.\ {\bf 75},  3340 (1995); A.\  Kumar, L.\
  Saminadayar,  D.C.\ Glattli, Y.\ Yin,  and B.\ Etienne, Phys.\ Rev.\
  Lett.\ {\bf 76}, 2778 (1996).
  
\bibitem{Feng} S.\  Feng,  P.A.\  Lee, and A.D.\  Stone,  Phys.\ Rev.\ 
  Lett.\ {\bf  56}, 1960 (1986);  A.A.\ Bobkov, V.I.\ Falko, and D.E.\ 
  Khmel'nitskii,  Zh.\ Eksp.\ Teor.\ Fiz.\  {\bf 98}, 703 (1990) [JETP
  {\bf 71},  393  (1990)];  see  also  S.\ Feng,  in  {\it  Mesoscopic
    Phenomena in Solids}, eds.\ B.L.\  Altshuler, P.A.\ Lee, and R.A.\ 
  Webb (Elsevier, Amsterdam, 1991).

\bibitem{Landauer} R.\  Landauer,    Phys.\ Rev.\ B   {\bf 47},  16427
  (1993).
  
\bibitem{AAK} B.L.\ Altshuler,  A.G.\ Aronov,  and D.E.\  Khmelnitsky,
  J.\ Phys.\   C, {\bf 15},   7367  (1982); see  also A.\  Stern,  Y.\ 
  Aharonov, and Y.\ Imry, Phys.\ Rev.\ B {\bf 41}, 3436 (1990).
  
\bibitem{Lee} B.L.\ Altshuler, JETP Lett.\ {\bf 41}, 648 (1985); P.A.\ 
  Lee,  A.D.\ Stone, and H.\ Fukuyama,  Phys.\  Rev.\ B {\bf 35}, 1039
  (1987).

\bibitem{Schmid}  A.\ Schmid, Z.\  Phys.\ {\bf 271}, 251 (1974); B.L.\ 
  Altshuler and A.\   G.\  Aronov, Sol.\  State Comm.\  {\bf  38},  11
  (1981).

\bibitem{details}   Details of  the   calculation   will be  published
  elsewhere, F.\ von Oppen and A.\ Stern (unpublished).

\bibitem{Rammer} For a  review, see  J.\  Rammer and H.\  Smith, Rev.\ 
  Mod.\ Phys.\ {\bf 58}, 323 (1986).

\bibitem{Larkin}  A.I.\ Larkin  and D.E.\  Khmel'nitskii, Sov.\ Phys.\ 
  JETP {\bf 64}, 1075 (1987).
  
\end{references}
\end{document}